\documentclass[12pt]{article} 
\usepackage{ol2}
\usepackage[draft]{hyperref}
\usepackage{amsmath}
\begin{document}
\title{Volume electric dipole origin of second-harmonic generation from metallic membrane with non-centrosymmetry patterns}
\author{Yong Zeng and Jerome V. Moloney}
\address{Arizona Center for Mathematical Sciences, University of
Arizona, Tucson, Arizona 85721}
\begin{abstract}
In this article, we analytically study second harmonic (SH)
generation from thin metallic films with subwavelength,
non-centrosymmetry patterns. Because the thickness of the film is
much smaller than the SH wavelength, retardation effects are
negligible. The far-field SH intensities are thus dominated by an
effective electric dipole. These analytical observations are further
justified numerically by studying the effect of polarization of the
fundamental field on both the SH signal and the electric dipole. It
is demonstrated that bulk SH polarization density is comparable with
its surface counterpart. The electric dipole, consequently,
originates from the entire \textit{volume} of the metallic membrane,
in contrast to the fact that SH generation from metal surface is
generally dominated by a \textit{surface} dipole.
\end{abstract}

\newpage
Optical second-harmonic generation (SHG) from one-dimensional metal
surfaces were first observed in 1965 \cite{brown}, and attracted
considerable attentions in the following fifty years (see
Ref.\cite{liebsch,yong2} and the cited references). Recently,
scientific interests have gradually shifted to the quadratic
nonlinearities of membrane-like metal films with subwavelength,
non-centrosymmetry patterns, partially owing to the significant
near-field enhancement induced by the excitation of localized
plasmon resonances \cite{busch,nahata,mcMahon}. On the experimental
side, SHG was observed from different geometric configurations such
as split-ring resonators \cite{klein,klein2} and their complementary
counterparts \cite{feth}, noncentrosymmetric T-shaped nanodimers
\cite{canfield}, T-shaped \cite{klein2} and L-shaped nanoparticles
\cite{kujala1,kujala2}.

For ideally infinite metal surfaces, it is well known that the
dominant second-harmonic (SH) electric dipole source appears only
at the interface between centrosymmetric media where the inversion
symmetry is broken. Higher order multipole sources, such as
magnetic dipole and electric quadrupole, merely provide a
relatively small bulk SH polarization density. In contrast to an
infinite surface, we will show in this article that there are
\textit{volume} electric dipoles which dominate SHG from thin
metal films with non-centrosymmetry patterns. The underlying
physical reasons include: (1) the non-centrosymmetry patterns
allow SH electric dipoles to appear in the overall structure
\cite{dadap2,finazzi}; and (2) the thickness of the film is far
smaller than the SH wavelength thus retardation effects are
negligible.

In general, SH radiation of a patterned structure satisfies the
following inhomogeneous wave equation \cite{jackson},
\begin{equation}
\nabla\times\nabla\times\mathbf{E}(\omega)-\frac{\omega^{2}}{c^{2}}\mathbf{E}=i\omega\mu_{0}f(\mathbf{r})\mathbf{j}(\mathbf{r},\omega),
\label{eq1}
\end{equation}
where $f(\mathbf{r})$ stresses the structural geometry: it equals
$1$ for metal and $0$ for vacuum. We now consider an ideal current
sheet whose thickness (along $z$ direction) $d_{z}\ll\lambda$ with
$\lambda$ being the SH wavelength. The SH electric field at
direction $\mathbf{n}$ in the far zone is found to be \cite{yong1}
\begin{equation}
\mathbf{E}=\frac{ik\eta}{4\pi
r}e^{ikr}(\mathbf{n}\times\mathbf{p})\times\mathbf{n}, \label{eq2}
\end{equation}
with $\eta=\sqrt{\mu_{0}/\epsilon_{0}}$ and
\begin{equation}
\mathbf{p}(\mathbf{n})=\int\int\mathbf{j}(\mathbf{r}_{\|},z)e^{-ik(\mathbf{n}_{\|}\cdot\mathbf{r}_{\|}+n_{z}z)}d\mathbf{r}_{\|}dz.
\label{eq3}
\end{equation}
Here $k=2\pi/\lambda$ is the wave number in vacuum, and the
integration is performed over the volume of the sheet. At either
positive or negative $z$ direction, we have
\begin{equation}
\mathbf{E}=\frac{ik\eta}{4\pi
r}e^{ikr}\mathbf{p}_{\|}(\mathbf{n}). \label{eq4}
\end{equation}
The far-zone field is therefore transverse with a vanishing $z$
component. Without loss of generality, we choose the electric
field along the $x$ direction (namely,
$\mathbf{p}_{\|}=p_{\|}\mathbf{e}_{x}$). The associated SH
intensity is thus proportional to
\begin{eqnarray}
&&\left|\int_{0}^{d_{z}}g(z)e^{-ikz}dz\right|^{2} \:\:\:\:
\textrm{at $+z$ direction};\cr
&&\left|\int_{0}^{d_{z}}g(z)e^{ikz}dz\right|^{2} \:\:\:\:\:\:
\textrm{at $-z$ direction}, \label{eq5}
\end{eqnarray}
where the integral function
\begin{equation}
g(z)=\int j_{x}(\mathbf{r}_{\|},z)d\mathbf{r}_{\|} \label{eq6}
\end{equation}
stands for the total current at a specific $z$ plane.

To identify the contributions of different order multipole
sources, we expand Eq.\eqref{eq5} in a power series of $kd_{z}$
\cite{jackson}. At the leading order, the intensities at the $\pm
z$ directions are identical and proportional to
\begin{equation}
\left|\int_{0}^{d_{z}}g(z)dz\right|^{2}=\left|\int\int
j_{x}(\mathbf{r}_{\|},z)d\mathbf{r}_{\|}dz\right|^{2}. \label{eq7}
\end{equation}
The right-hand side of Eq.\eqref{eq7} is the total current of the
sheet or equivalently the effective \textit{electric dipole}. At
the first order of $kd_{z}$, these two signals have a difference
of
\begin{equation}
2ik\int_{0}^{d_{z}}\int_{0}^{d_{z}}
g(z_{1})g^{\ast}(z_{2})(z_{1}-z_{2})dz_{1}dz_{2}, \label{eq8}
\end{equation}
and the associated relative difference reads as
\begin{equation}
\text{Im}\left\{\frac{-4k\int_{0}^{d_{z}}
g(z)zdz}{\int_{0}^{d_{z}}g(z)dz}\right\}. \label{eq9}
\end{equation}
In other words, the difference is induced by the appearance of
magnetic dipole and electric quadrupole. Their influence is
negligible under the conditions such as (1) complex $g(z)$ with
slowly-varying phase, or (2) $d_{z}/\lambda\ll 1$, or (3) $g(z)$ is
symmetrical such that $g(z)=g(d_{z}-z)$; or (4) $g(z)$ is localized
at one point (such as SHG from metal surface). On the other hand,
$g(z)$ with rapid-varying phase, such as a nonlocal electric dipole,
can lead to a remarkable difference \cite{dadap2,dadap,bachelier}.
However, we will show it is unlikely to happen in an ideally thin
film.

To numerically study second-order nonlinearities of thin metallic
films, we employ a classical model developed recently
\cite{yong2,feth}. This model has been demonstrated to not only
provide qualitative agreement with experiments, but also reproduce
the overall strength of the experimentally observed SH signals.
Before we present the numerical results, it should be stressed that,
as suggested by simulations (not shown here), these results are
quite general and can be found from different configurations
including (1) perfect T-shaped, C-shaped and L-shaped patterns; (2)
ideal $C_{1v}$-symmetric patterns supported by semi-infinite
dielectric substrate; (3) asymmetric pattern with perturbed
$C_{1v}$-symmetry; and (4) particle arrays with size-tolerance
disorder.

\begin{figure}[t]
\centering
\includegraphics[width=0.6\textwidth]{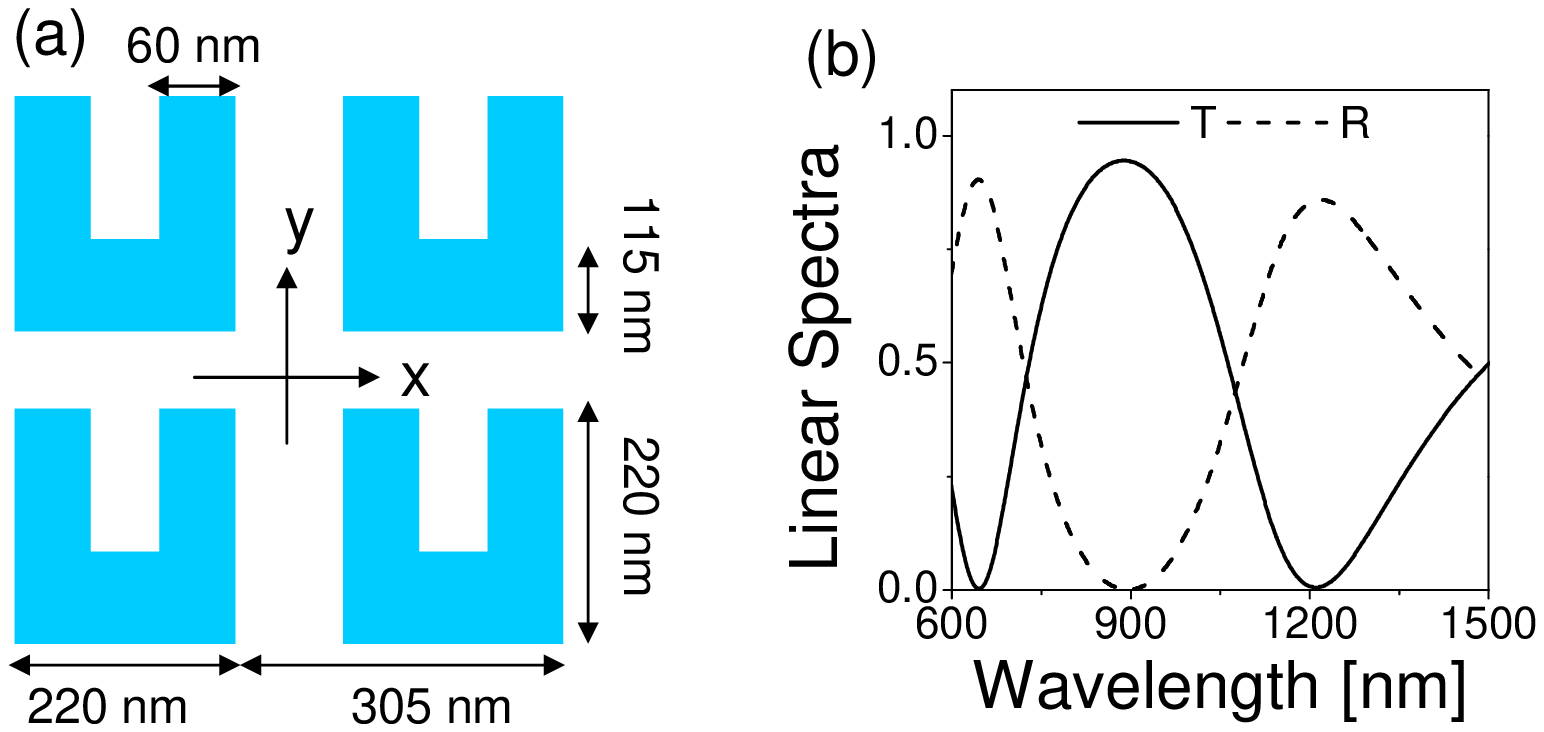}\vspace*{-6.0cm}
\caption{(a) Schematic drawing of a square lattice of gold
split-ring resonators. The thickness of the gold film is 25 nm.
(b) Its linear spectra under a $x$-polarized normal incidence.}
\label{fig2}
\end{figure}

The representative film studied here is a freestanding array of
gold split-ring resonators with a thickness of 25 nm, which
closely matches the experimental sample in
Ref.\cite{klein,klein2,feth}. Schematic drawing of this patterned
film is shown in Figure \ref{fig2}, together with its linear
spectra. In the following, we choose a fundamental-frequency (FF)
field which propagates along the $z$ direction and has a
wavelength of 1212 nm. This wavelength corresponds to the valley
in the transmission which is induced by the fundamental plasmonic
resonance \cite{rockstuhl}.

\begin{figure}
\centering
\includegraphics[width=0.6\textwidth]{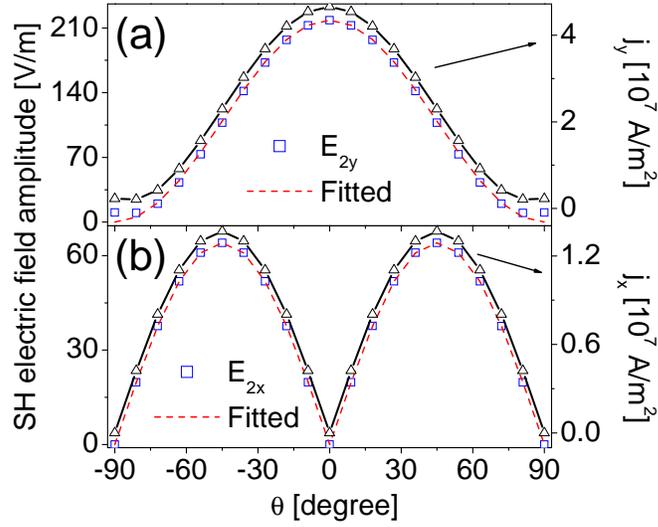}\vspace*{-4.5cm}
\caption{The effect of incident polarization angle $\theta$ on the
far-zone second harmonic electric field along the $\pm z$
direction, with $\theta=0$ corresponding to polarization along the
$x$ direction. The fitted function employed is (a)
$\cos^{2}\theta$ and (b) $|\sin2\theta|$, respectively. The
averaged current density (proportional to the electric dipole)
$\mathbf{j}$ is also shown. The amplitude of the incident
fundamental frequency electric field $E_{0}$ is $2\times
10^{7}$(V/m), same as that used in the experiments \cite{klein}.
The bulk plasma frequency of gold is taken as
$\omega_{p}=1.367\times 10^{16} s^{-1}$, the phenomenological
collision frequency $\gamma=6.478\times 10^{13} s^{-1}$
\cite{yong2}.} \label{fig3}
\end{figure}

Figure \ref{fig3} shows the effect of the incident polarization
angle $\theta$ on both the far-zone SH electric field as well as the
total SH current. The distinct dependence between the different
component of the SH electric field can be interpreted in terms of
the electric dipole approximation \cite{yong2}. The most striking
feature is that almost identical SH electric fields, with a relative
difference as small as 0.5\%, are found along the ($\pm z$)
directions. Moreover, the amplitude of the electric dipole is
proportional to the SH electric fields. Consequently, SH radiation
from this thin film is dominated by the electric dipole defined in
Eq.(\ref{eq7}), and higher-order multipoles can be safely neglected.
It should be mentioned that the same observations are also
experimentally found from an individual metal tip: the emission of
SH radiation at the tip can be attributed to a single on-axis
oscillating dipole \cite{bouhelier,nappa}.

\begin{figure}[t]
\centering
\includegraphics[width=0.8\textwidth]{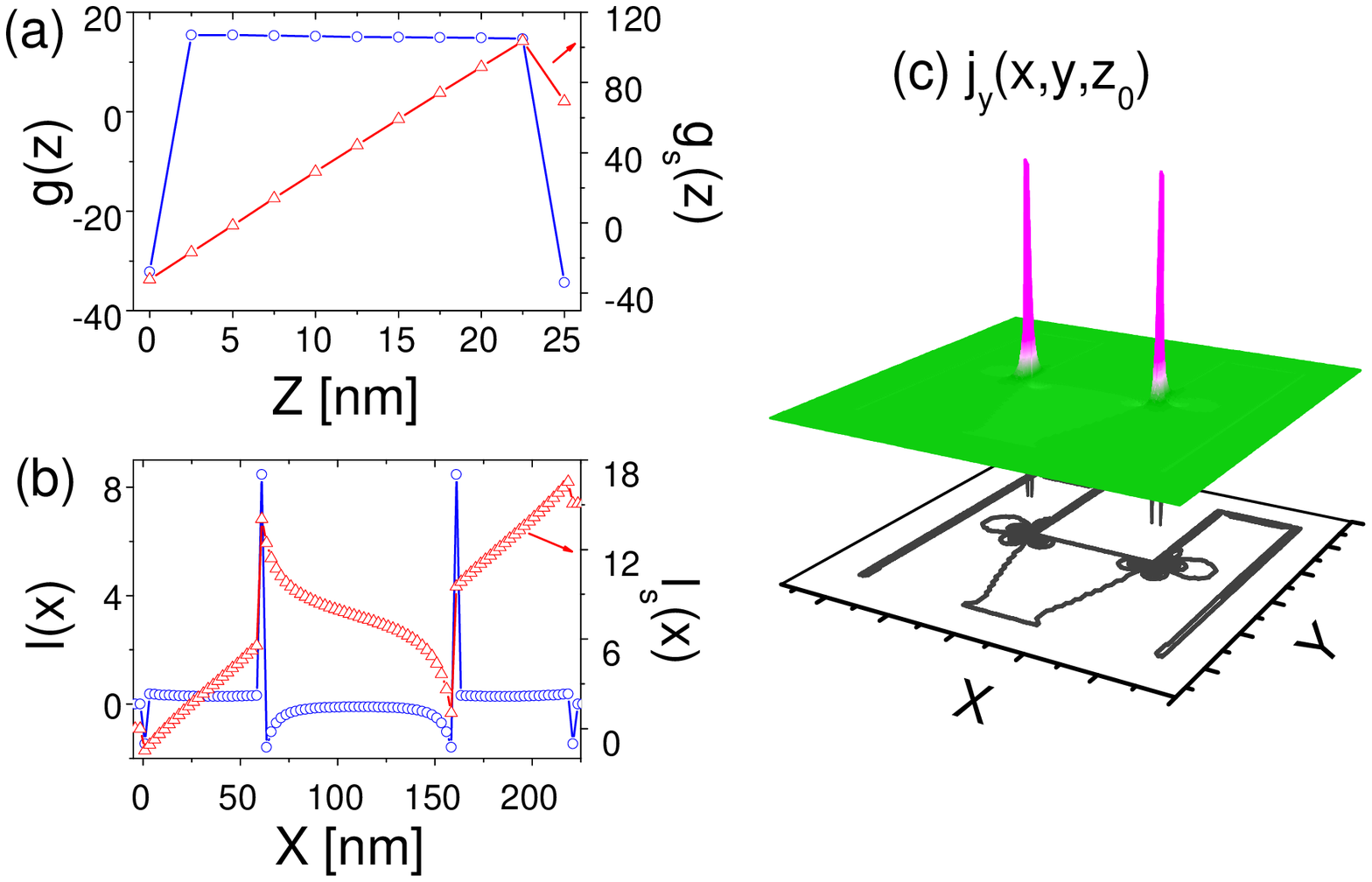}\vspace*{-5.0cm}
\caption{The spatial distribution of second-harmonic current (with
arbitrary units) with a $x$-polarized normal incidence. The
resolution is limited to 2.5 nm due to computational resource
limitations. $z(x)=0$ corresponds to the starting plane of the
gold film along the $z(x)$ direction. (c) The current distribution
on the plane of $z_{0}=12.5$ nm.} \label{fig4}
\end{figure}

To explore the characteristics of the electric dipole, the spatial
distribution of $j_{y}$ with a $x$-polarized illumination is
calculated ($\theta=0$ therefore $j_{x}=0$) and plotted in Figure
\ref{fig4}, at a time corresponding to a maximal $j_{y}$. We
further define the following functions
\begin{eqnarray}
&&g(z)=\sum_{x,y}j_{y}(x,y,z),\:\:\:\: g_{s}(z)=\sum_{z'\leq
z}g(z');\cr &&l(x)=\sum_{y}j_{y}(x,y,z_{0}),\:\:\:\:
l_{s}(x)=\sum_{x'\leq x}l(x'); \label{eq9}
\end{eqnarray}
with $z_{0}=12.5$ nm. Here $g(z')$ stands for the total current at
the $z=z'$ plane, and $l(x')$ represents the sum of $j_{y}$ along
the line $(x=x', z=z_{0})$. $g_{s}(z)$ is a cumulative function with
$g_{s}(z_{max})$ corresponding to the electric dipole. Clearly, the
$g(z)$ here is almost symmetrical in $z$, mainly because the film
thickness is so thin that both FF and SH fields are nearly constant
inside the film. In other words, the retardation effects are
negligible. As discussed previously, it is this symmetrical $g(z)$
which results in the negligible higher-order multipoles. On the
other hand, each element of the metal volume contributes to the SH
radiation, in a constructive/destructive way. Assuming the outermost
grid layer (a finite-difference time-domain method is utilized here
\cite{yong2}) constitutes the metal surface (its 2.5-nm thickness is
much thicker than a \textit{real} surface), then the internal bulk
polarization density is found to be comparable with its surface
counterpart. As a consequence, the effective electric dipole
originates from the entire metal \textit{volume}. In addition, SH
current distribution on the $z=12.5$ nm plane is shown in Fig.
(\ref{fig4}c). Around the inner corners, the current is found to be
strongly depolarized and form localized bright hot spots. This
phenomenon has been observed from metal surfaces with nanoscale
roughness, and the origin is attributed to the overlap between SH
and fundamental modes \cite{anceau,stockman}.

We now discuss the experiment reported in Ref.\cite{kujala1} where
considerable multipole interferences, in addition to the dominant
electric dipole, were observed. The experimental sample is a
20-nm-thick gold film periodically perforated with an L-shaped
pattern, and the FF wavelength is 1060 nm. The phase variation of
the FF wave inside the metal is small, approximately $0.01\pi$,
which should result in a $g(z)$ quite close to the one shown in Fig.
(\ref{fig4}a). Consequently, the multipole contribution should be
negligible. The difference between experiment and theory is believed
to be induced by sample imperfections which may result in strongly
localized out-of-phase currents at \textit{both} (according to the
condition (4) mentioned above) top and bottom metal-dielectric
interfaces. On the other hand, we want to point out that significant
multipole sources may appear in large-size metal particles
\cite{dadap2,dadap} and multilayered structures such as bulk
photonic metamaterials \cite{kim}.

The authors thank Stephan W. Koch and Colm Dineen for their helpful
comments. This work is supported by the Air Force Office of
Scientific Research (AFOSR), under Grant No. FA9550-07-1-0010 and
FA9550-04-1-0213. J. V. Moloney acknowledges support from the
Alexander von Humboldt foundation.

\newpage

\newpage


\newpage
\begin{thebibliography}{99}
\bibitem{brown} F. Brown, R. E. Parks, A. M. Sleeper, ``Nonlinear Optical Reflection from a Metallic
Boundary", Phys. Rev. Lett. 14, 1029 (1965).
\bibitem{liebsch} A. Liebsch, {\it Electronic Excitations at Metal Surfaces} (Plenum press, 1997).
\bibitem{yong2} Y. Zeng, W. Hoyer, J. Liu, S. W. Koch, J. V.
Moloney, ``A classical theory for second-harmonic generation from
metallic nanoparticles", Submitted, also at arXiv:0807.3575.
\bibitem{busch} K. Busch, G. von Freymann, S. Linden, S. F. Mingalaeev, L. Tkeshelashvili, and M.
Wegener, ``Periodic nanostructures for photonics", Phys. Rep. 444,
101 (2007).
\bibitem{nahata} A. Nahata, R. A. Linke, T. Ishi, K. Ohashi, ``Enhanced nonlinear optical conversion from a periodically nanostructured metal
film", Opt. Lett. 28, 423 (2003).
\bibitem{mcMahon} M. D. McMahon, R. Lopez, R. F. Haglund, E. A. Ray, P. H.
Bunton, ``Second-harmonic generation from arrays of symmetric gold
nanoparticles", Phys. Rev. B 73, 041401(R) (2006).
\bibitem{klein} M. W. Klein, C. Enkrich, M. Wegener, S. Linden, ``Second-Harmonic Generation from Magnetic Metamaterials," Science 313, 502
(2006).
\bibitem{klein2} M. W. Klein, M. Wegener, N. Feth, S. Linden,
``Experiments on second- and third-harmonic generation from
magnetic metamaterials", Opt. Express 15, 5238 (2007).
\bibitem{feth} N. Feth, S. Linden, M. W. Klein, M. Decker, F.
Niesler, Y. Zeng, W. Hoyer, J. Liu, S. W. Koch, J. V. Moloney, and
M. Wegener, ``Second-harmonic generation from complementary
split-ring resonators", Opt. Lett. 33, 1975 (2008).
\bibitem{canfield} B. K. Canfield, H. Husu, J. Laukkanen, B. Bai,
M. Kuittinen, J. Turunen, M. Kauranen, ``Local field asymmetry
drives second-harmonic generation in noncentrosymmetric
nanodimers," Nano Lett. 7, 1251 (2007).
\bibitem{kujala1} S. Kujala, B. K. Canfield, M. Kauranen, Y. Svirko, J. Turunen, ``Multipole Interference in the Second-Harmonic Optical Radiation from Gold Nanoparticles'', Phys. Rev. Lett. 98, 167403 (2007).
\bibitem{kujala2} S. Kujala, B. K. Canfield, M. Kauranen, Y. Svirko, J. Turunen, ``Multipolar analysis of second-harmonic radiation from gold nanoparticles'', Opt. Express, 16, 17196
(2008).
\bibitem{dadap2} J. I. Dadap, J. Shan, T. F.
Heinz, ``Theory of optical second-harmonic generation from a
sphere of centrosymmetric material: small-particle limit", J. Opt.
Soc. Am. B 21, 1328 (2004).
\bibitem{finazzi} M. Finazzi, P. Biagiona, M. Celebrano, L.
Du\`{o}, ``Selection rules for second-harmonic generation in
nanoparticles", Phys. Rev. B 76, 125414 (2007).
\bibitem{jackson} J. D. Jackson, {\it Classical Electrodynamics} (Second edition, John Wiley \& Sons, 1975).
\bibitem{yong1} Y. Zeng and J. V. Moloney, "Polarization-Current-Based, finite-difference time-domain, Near-to-Far-Field
Transformation", Opt. Lett. In Press (2009), also at
arXiv:0903.0663.
\bibitem{dadap} J. I. Dadap, J. Shan, K. B. Eisenthal, T. F.
Heinz, ``Second-Harmonic Rayleigh Scattering from a Sphere of
Centrosymmetric Material", Phys. Rev. Lett. 83, 4045 (1999).
\bibitem{bachelier} G. Bachelier, I. Russier-Antoine, E. Benichou, C. Jonin, P. F.
Brevet, ``Multipolar second-harmonic generation in noble metal
nanoparticles", J. Opt. Soc. Am. B 25, 955 (2008).
\bibitem{rockstuhl} C. Rockstuhl, F. Lederer, C. Etrich, Th. Zentgraf, J. Kuhl, and H.
Giessen, ``On the reinterpretation of resonances in
split-ring-resonators at normal incidence", Opt. Express 14, 8827
(2006).
\bibitem{bouhelier} A. Bouhelier, M. Beversluis, A. Hartschuh, and L.
Novotny, ``Near-Field Second-Harmonic Generation Induced by Local
Field Enhancement'', Phys. Rev. Lett. 90, 013903 (2003).
\bibitem{nappa} J. Nappa, G. Revillod, I. Russier-Antoine, E.
Benichou, C. Jonin, P. F. Brevet, ``Electric dipole origin of the
second harmonic generation of small metallic particles'', Phys.
Rev. B 71, 165407 (2005).
\bibitem{anceau} C. Anceau, S. Brasselet, J. Zyss, P. Gadenne, ``Local second-harmonic generation enhancement on gold nanostructures probed by two-photon
microscopy", Opt. Lett. 28, 713 (2003).
\bibitem{stockman} M. I. Stockman, D. J. Bergman, C. Anceau, S. Brasselet, J.
Zyss, ``Enhanced Second-Harmonic Generation by Metal Surfaces with
Nanoscale Roughness: Nanoscale Dephasing, Depolarization, and
Correlations", Phys. Rev. Lett. 92, 057402 (2004).
\bibitem{kim} E. Kim, F. Wang, W. Wu, Z. Yu, Y. R. Shen, ``Nonlinear optical spectroscopy of photonic metamaterials", Phys. Rev. B
78, 113102 (2008).
\end{thebibliography}
\end{document}